\RequirePackage{fix-cm}
\documentclass[smallcondensed]{svjour3}     
\smartqed  
\usepackage{graphicx}
\usepackage{mathptmx}      
\newcommand{\s}{\mbox{$\sigma_1$}}
\newcommand{\p}{\mbox{$\pi_1$}}
\newcommand{\pbar}{\mbox{$\overline{\mathrm p}$}}
\newcommand{\Hbar}{\mbox{$\overline{\mathrm H}$}}

 \journalname{Hyperfine Interactions}
\begin{document}

\title{Hyperfine spectroscopy of hydrogen and antihydrogen in ASACUSA}


\author{{E. Widmann} \and
		C.~Amsler\and
        S.~Arguedas Cuendis\and
		H.~Breuker\and
	 	M.~Diermaier\and
		P.~Dupr\'e\and
        C.~Evans\and
		M.~Fleck\and
		A.~Gligorova\and
		H.~Higaki\and
		Y.~Kanai\and
        B.~Kolbinger\and
        N.~Kuroda\and
		M.~Leali\and
        A.M.M.~Leite\and
		V.~M\"ackel\and
        C.~Malbrunot\and
		V.~Mascagna\and
        O.~Massiczek\and
		Y.~Matsuda\and
		D.J.~Murtagh\and
		Y.~Nagata\and
        A.~Nanda\and
        D.~Phan\and
        C.~Sauerzopf\and
        M.C.~Simon\and
        M.~Tajima\and
        H.~Spitzer\and 
        M.~Strube\and
        S.~Ulmer\and  
        L.~Venturelli\and
        M.~Wiesinger\and 
        Y.~Yamazaki\and
        J.~Zmeskal
}



\institute{E. Widmann, C. Amsler, S. Arguedas Cuendis, M. Diermaier, A. Gligorova, B. Kolbinger, A.M.M. Leite, C. Malbrunot, O.Massiczek, D.J. Murtagh, A. Nanda, D. Phan, C. Sauerzopf, M.C. Simon, H. Spitzer, M. Strube, M. Wiesinger, J. Zmeskal  \at
              Stefan Meyer Institute for Subatomic Physics, Austrian Academy of Sciences, 1090 Vienna, Austria \\
            \email{eberhard.widmann@oeaw.ac.at}             \\
            \emph{Present address:} of S. Arguedas Cuendis: CERN, Geneva, Switzerland \\            \ 
            \emph{Present address:} of M. Wiesinger: Max Planck Institute for Nuclear Physics, Heidelberg, Germany
           \and
           C. Malbrunot \at
              CERN, Geneva Switzerland   
           \and 
           H. Breuker, P.~Dupr\'e, Y. Kanai, V. M\"ackel, M. Tajima, S. Ulmer, Y. Yamazaki \at
           RIKEN, 351-0198 Saitama, Japan 
           \and 
           N. Kuroda, M. Fleck, Y. Matsuda \at
           University of Tokyo, 153-8902 Tokyo, Japan   \\
           \emph{Present address:} of M. Tajima: RIKEN, 351-0198 Saitama, Japan 
           \and 
           H. Higaki \at
           Hiroshima University, 739-8530 Hiroshima, Japan
           \and 
           C. Evans, M. Leali, V. Mascagna, L. Venturelli \at
           Dipartimento di Ingegneria dell'Informazione, Universit\`a' degli Studi di Brescia, 
           Brescia, Italy and Istituto Nazionale di Fisica Nucleare (INFN), sez. Pavia, Italy \\
           \emph{Present address:} of V. Mascagna: Dipartimento di Scienza e Alta Tecnologia, Univerit\`a' dell'Insubria and  INFN sez. di Pavia, Italy
           \and 
           Y. Nagata \at
           Tokyo University of Science, 162-8601 Tokyo, Japan
}

\date{Received: date / Accepted: date}

\maketitle

\begin{abstract}
The ASACUSA collaboration at the Antiproton Decelerator of CERN aims at a precise measurement of the antihydrogen ground-state hyperfine structure as a test of the fundamental CPT symmetry. A beam of antihydrogen atoms is formed in a CUSP trap, undergoes Rabi-type spectroscopy  and is detected downstream in a dedicated antihydrogen detector. In parallel measurements using a polarized hydrogen beam are being performed to commission the spectroscopy apparatus and to perform measurements of parameters of the Standard Model Extension (SME). The current status of antihydrogen spectroscopy is reviewed and progress of ASACUSA is presented.
\end{abstract}

\section{Antihydrogen and CPT symmetry}
\label{intro}

CPT symmetry is a cornerstone of the Standard Model (SM) of particle physics due the existence of the CPT theorem
which states that in any Lorentz-invariant, unitary, local field theory of point-like particles is invariant under the combined operation of charge conjugation C, parity P and time reversal T. The observed matter - antimatter asymmetry in the Universe as well as theories beyond the SM like string theory, where some of the prerequisites of the mathematical proof of the CPT theorem do not hold any more, have encouraged experimental tests of CPT symmetry.

\begin{figure}[b]
\begin{center}
\includegraphics[width=0.6\textwidth]{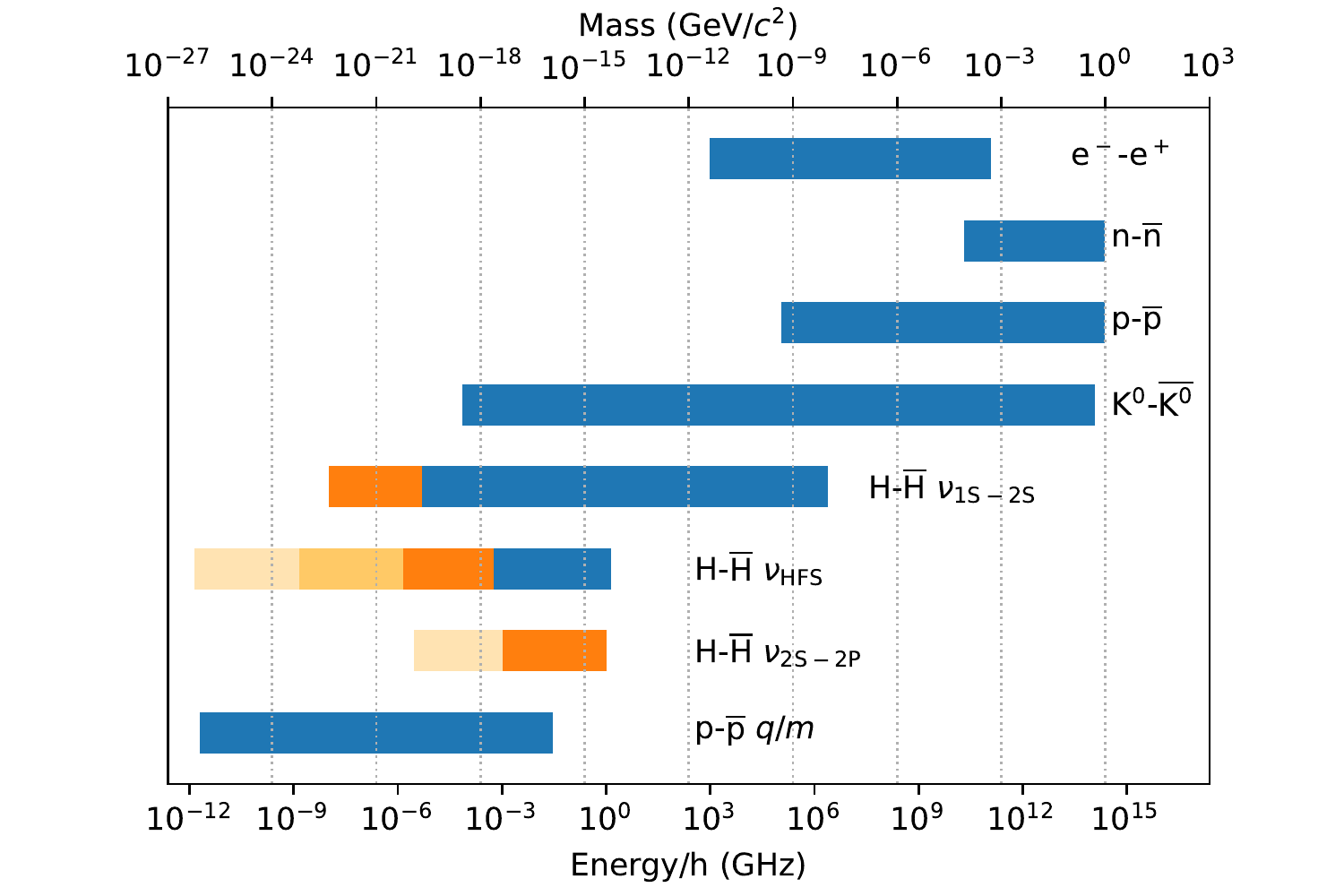}
\caption{Comparison of several tests of CPT symmetry on an energy scale. Bar's right hand side: measured quantity, length of bar: relative precision of CPT test, left hand side: sensitivity on an absolute energy scale. Blue: existing test. Orange: predicted sensitivity if existing precision for hydrogen is achieved. In the case of HFS: orange: first goal for in-beam measurement, paler orange: line width for fountain, yellow: hydrogen maser result. In case of Lamb shift: orange: estimated achievable accuracy, paler orange: accuracy for hydrogen. Values are from PDG \cite{TanabashiEtAl2018} except for \Hbar\ results for HFS \cite{Ahmadi2017}, 1S--2S \cite{Ahmadi2018}, and Lamb shift \cite{Crivelli2016}. \pbar\ charge-to-mass ratio \cite{Ulmer:2015}, note that in this case only the left hand side is well defined, while the right hand side is less precise since the cyclotron frequency is proportional to the magnetic field in which the measurement is taken.}
\label{fig:CPT}       
\end{center}
\end{figure}

One of the consequences of the CPT theorem is that particles and antiparticles have exactly the same (mass, total lifetime) or exactly the opposite (charge, magnetic moment) properties. An overview of existing measurements is available at the Particle Data Group \cite{TanabashiEtAl2018}, some of the standard measurements of masses of elementary particles and antiparticles are shown in Fig.~\ref{fig:CPT} (a more elaborate comparison including e.g. $g$-factor measurements as performed by the BASE collaboration to ppb precision \cite{Smorra2017} is possible within the Standard Model Extension (SME) framework \cite{Colladay:1997vn,Kostelecky:2011jr,Kostelecky:2018}).

From Fig.~\ref{fig:CPT} it becomes evident that atomic physics measurements like the cyclotron frequency of protons and antiprotons in a trap yielding the charge-to-mass ratio $q/m$ and determinations of the internal structure of antihydrogen offer the most sensitive tests of CPT symmetry on an absolute scale. The potential of antihydrogen, the simplest anti-atom whose matter equivalent hydrogen is one of the best studied systems in physics, for precision CPT tests has been realized early  \cite{CharltonEtAl1994}, the most difficult challenge being the formation of cold antihydrogen atoms suitable for spectroscopy. At the Antiproton Decelerator of CERN now several experiments are routinely producing antihydrogen \cite{HORI2013206}, with the ALPHA collaboration having been able to perform first spectroscopy measurements recently \cite{Ahmadi2017,Ahmadi2018,Ahmadi2018a}. 

ASACUSA is aiming at a measurement of the antihydrogen ground-state hyperfine structure GS-HFS \cite{Widmann:2001fk,Widmann:2013qy}, transporting the \Hbar\ atoms to a region far away from the stray magnetic fields at the formation region, like it was done for hydrogen by Rabi \cite{Rabi:1938tq} and others. This will ultimately allow for precision exceeding the one of in-trap hyperfine spectroscopy as done in \cite{Ahmadi2018}. 

\begin{figure}[b]
  \raisebox{-0.5\height}{\includegraphics[width=0.6\textwidth]{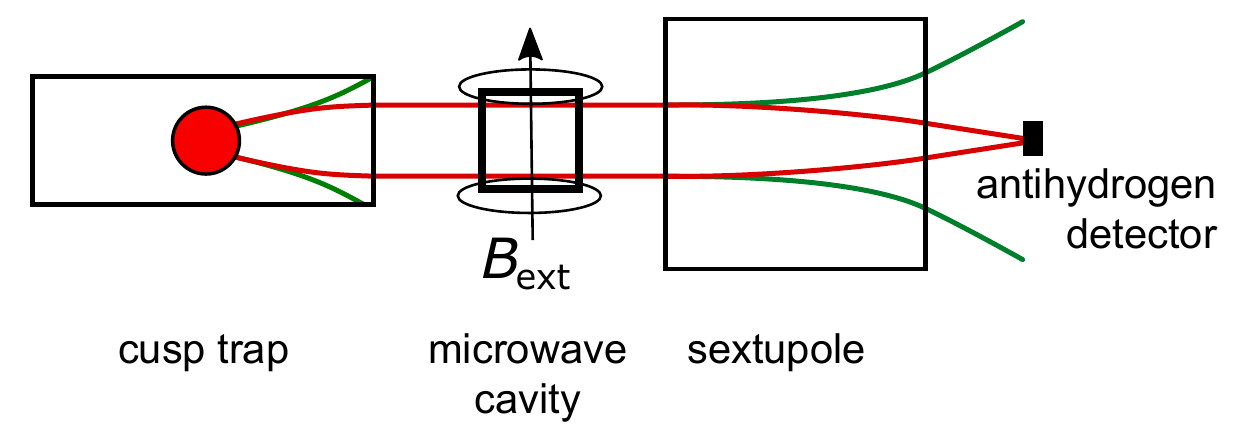}}
  \raisebox{-0.5\height}{\includegraphics[width=0.4\textwidth]{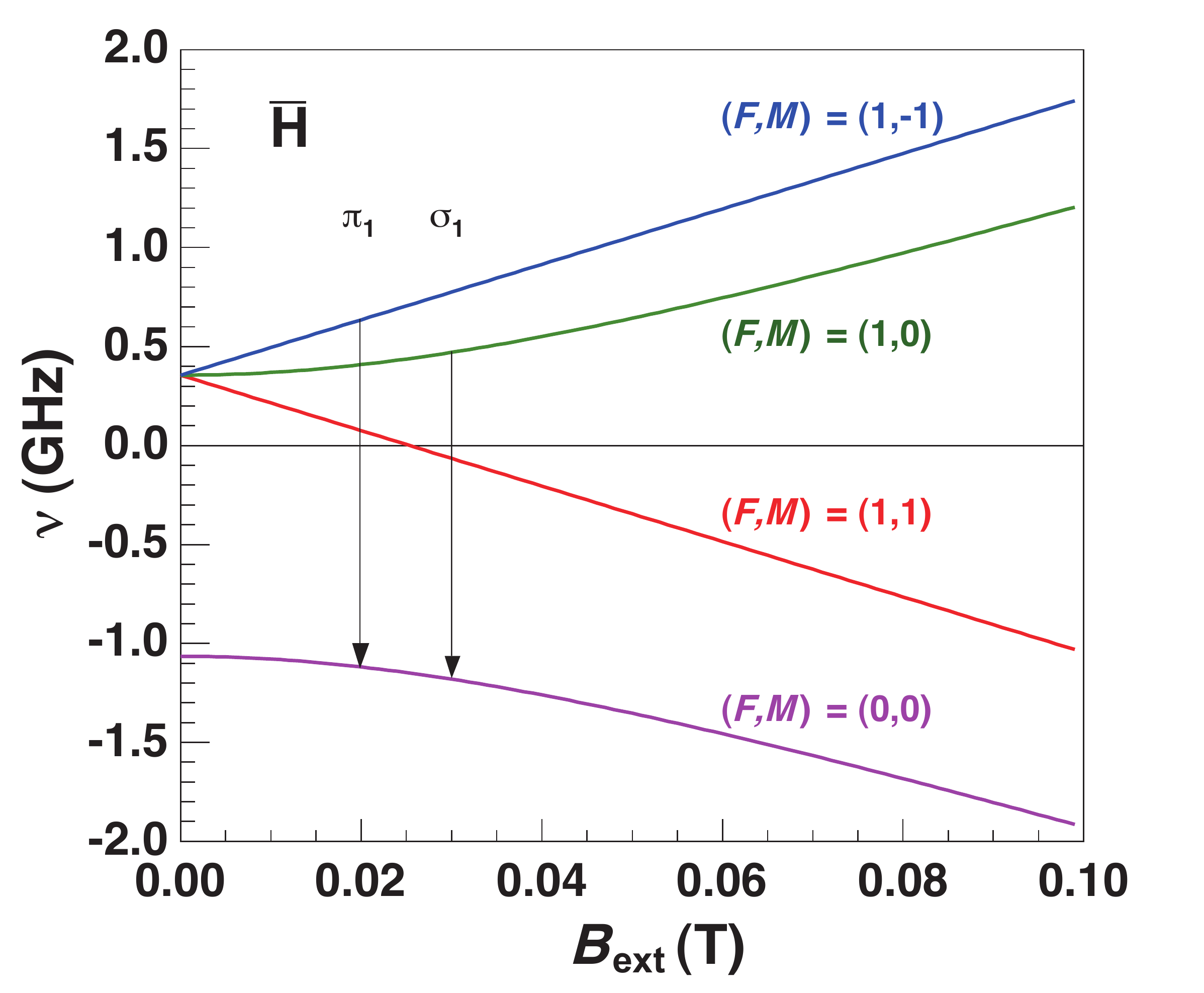}}
\caption{Left: schematic antihydrogen formation and hyperfine measurement beam line. Antihydrogen is created by mixing antiprotons and positrons in the CUSP trap that polarizes the outgoing beam by focusing low-field seekers (red) and defocussing high-field seekers (green). The beam then passes a microwave cavity where spin-flips are induced under the presence of an external static magnetic field $B_\mathrm{ext}$, a superconducting sextupole which again selects low-field seekers, and an antihydrogen detector. Right: Breit-Rabi diagram showing the magnetic field dependence of the four hyperfine states of antihydrogen.}
\label{fig:Setup}       
\end{figure}

\section{Antihydrogen beam formation and hyperfine spectroscopy}
\label{sec:Hbar}

In ASACUSA antihydrogen atoms are formed from their ingredients \cite{doi:10.7566/JPSCP.18.011009}. Antiprotons from the CERN Antiproton Decelerator are trapped and cooled in a  Penning trap MUSASHI \cite{Kuroda:2005ik}, Positrons are produced by a $^{22}$Na source in combination with a solid Ne ice moderator, they are accumulated in a Surko type buffer gas trap. Both species are brought together and mixed in a magnetic CUSP configuration \cite{Mohri:2003wu} using a nested Penning trap configuration \cite{Gabrielse1988} which is widely used at the AD.

Fig.~\ref{fig:Setup} (left) shows a schematic view of the Rabi-type beam spectroscopy experiment. 
Fig.~\ref{fig:Setup} (right) depicts the behavior of the four hyperfine states characterized by the  quantum number $F$ connected to the total spin $\vec{F}=\vec{S_1}+\vec{S_2}$ and its projection $M$ to the quantization axis. States with increasing (decreasing) energy in an external magnetic field are called low-field seekers (high-field seekers). Two possible transitions induced by microwaves (\s, \p) that are accessible in this experiment are denoted by arrows. 

After the initial success of \Hbar\ formation \cite{Enomoto:2010uq}, a beam of cold antihydrogen atoms was observed 2.7 m downstream of the formation region \cite{Kuroda:2014fk}. In 2017 for the first time the principle quantum number distribution could be measured using the antihydrogen detector \cite{Kolbinger2018} and a field ionization stage, which -- as expected from three-body formation -- contains mainly Rydberg atoms \cite{Malbrunot2018}. \Hbar\ with $n<14$ were detected with 4.5 $\sigma$ significance, which could within their life time of $\sim 100 \, \mu$s decay to the ground state before reaching the cavity. The overall observed rate of such events of $\sim 0.001$ s$^{-1}$ is however far too small for spectroscopy measurements. The primary task is to increase the production rate of ground-state antihydrogen by 1--2 orders of magnitude.


\section{Hydrogen hyperfine spectroscopy results}
\label{sec:H}

Due to the scarce availability of antihydrogen, ASACUSA decided to develop a hydrogen beam with the initial goal to commission the hyperfine spectroscopy apparatus. A source of cold polarized hydrogen with a temperature of $\sim50$ K as expected for the \Hbar\ formed in the CUSP was connected to the microwave cavity and superconducting sextupole magnet to be used for antihydrogen spectroscopy. For hydrogen detection a commercial Q-mass spectrometer was used. With this setup the \s\ transition in hydrogen could be determined to a few Hertz precision at $\nu \sim 1.42$ GHz, resulting in a 2.7 ppb measurement of the hydrogen GS-HFS \cite{Diermaier2017}. This constitutes the most precise determination of this quantity in a beam so far, the highest precision of 2 mHz having been obtained in a hydrogen maser \cite{Hellwig:1970,Karshenboim2000}), which cannot be used for antimatter. The result using the \s\ transition also allows us to estimate that with this method about 8000 \Hbar\ atoms in the ground state are needed to obtain a precision of $\sim 1$  ppm, which is the initial goal of ASACUSA.

\begin{figure}[b]
  \includegraphics[width=1.\textwidth]{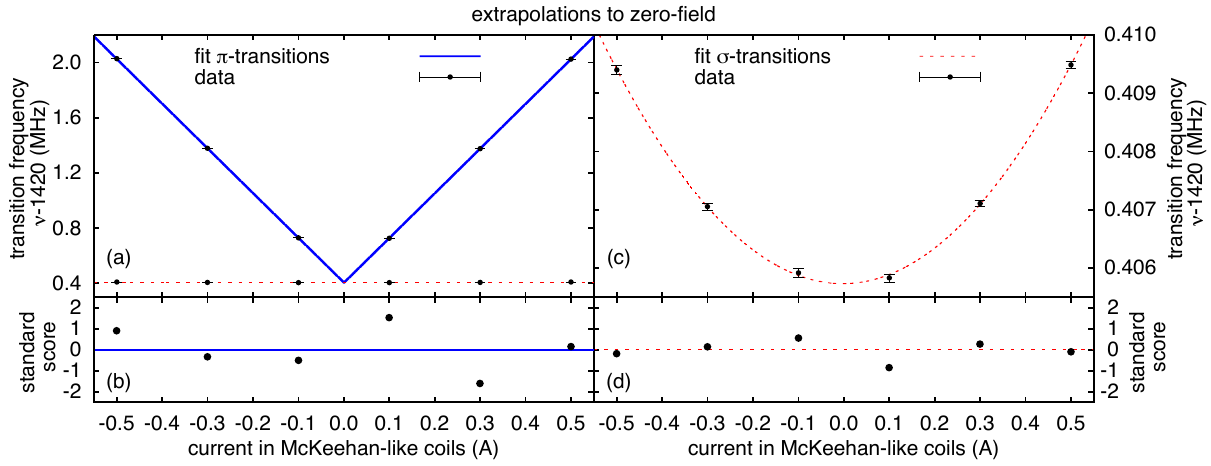}
\caption{Preliminary results of the first simultaneously taken extrapolations of $\nu_\pi$ (a) and $\nu_\sigma$ (c) as a function of the external static magnetic field for hydrogen. (b) and (d) show the deviations from the fit as standard scores. The dashed red line in panel (a) shows $\nu_\sigma$ to illustrate the much smaller $B$-field dependence compared to $\nu_\pi$.}
\label{fig:hextrapol}       
\end{figure}

While many coefficients of the minimal SME have been constrained by hydrogen maser measurements, in the newly developed non-minimal SME a set of coefficients is present that has not yet been investigated experimentally \cite{KosteleckyVargas2015}, especially those depending on the orientation of the static magnetic field with respect to the Earth rotation axis \cite{Malbrunot2018}. Within the SME, the \p\ transition is sensitive to CPT violations while the \s\ transition is not. Since the \p\ transition is more sensitive to magnetic field inhomogeneities, we constructed new McKeehan-like coils \cite{McKeehan1936} and a three-layer cylindrical shielding \cite{ArguedasCuendis2017}. Additionally we changed the beam optics so to produce the same rate of hydrogen atoms in the different quantum states at the detector \cite{Wiesinger2017}, using permanent sextupole magnets and ring apertures. With this setup, first measurements were done determining alternatingly the transition frequencies $\nu_\sigma$ and $\nu_\pi$ at each setting of the external constant magnetic field $B_\mathrm{ext}$, which is controlled by the McKeehan-like coils. Fig.~\ref{fig:hextrapol} shows preliminary measurements of $\nu_\sigma$ and $\nu_\pi$ at several $B_\mathrm{ext}$. The $B$-field dependence of the two transitions follows from the Breit-Rabi formula \cite{BreitEtAl1931}, yielding a hyperbolic dependence of $\nu_\sigma (B_\mathrm{ext})$ and a dominantly linear one for $\nu_\pi (B_\mathrm{ext})$. Table~\ref{tab:Hext} gives the values of the hyperfine transition frequency $\nu_0$ obtained by extrapolating each type of transition to zero external field, yielding typical precisions of a 10--30 ppb in a first test run lasting five days.

Instead of extrapolations, the measurement of $\nu_\sigma$ and $\nu_\pi$ at the same $B_\mathrm{ext}$ can be used to directly determine the zero-field GS-HFS frequency $\nu_0$ \cite{KolbingerEtAl2015}. This method, using the same data sets, leads to a factor 2 smaller errors than the individual extrapolations and is of interest to antihydrogen, as it includes the more interesting \p\ transition and requires the acquisition of only two resonances to predict the zero-field value (thereby reducing the  amount of required \Hbar\ events to 4000 for a $\sim 1$ ppm result). Further tests of this method are ongoing and systematic measurements of both $\nu_\sigma$ and $\nu_\pi$ are planned in the near future.


\begin{table}[h]
\caption{Hydrogen ground-state hyperfine splitting results using the ASACUSA hydrogen beam (preliminary). The results agree within 1 standard deviation to the literature value $\nu_\mathrm{lit}$ obtained by a hydrogen maser \cite{Hellwig:1970}. From \cite{ArguedasCuendis2017}.} \label{tab:Hext}
\begin{tabular}{lccc}
\hline\noalign{\smallskip}
                                           & $\nu_0$ {[}Hz{]}    & Relative error  & $\nu_0 - \nu_\mathrm{lit}$ {[}Hz{]}       \\ \hline\noalign{\smallskip}
                                           
$\sigma_1$ extrapolation                                                                        & 1~420~405~767(15) & 1.04 $\times 10^{-8}$                                                                                & 15 \\ \hline\noalign{\smallskip}
$\pi_1$ extrapolation                                                                           & 1~420~405~760(34) & 2.38 $\times 10^{-8}$                                                                                 & 8 \\ \hline\noalign{\smallskip}
\begin{tabular}[l]{@{}l@{}}Mean value of the two \\ extrapolations\end{tabular}                   & 1~420~405~766(14) & 9.96 $\times 10^{-9}$                                                                                 & 14 \\ \hline\noalign{\smallskip}
\begin{tabular}[l]{@{}l@{}}$\nu_\sigma$ and $\nu_\pi$ determined \\ at same static magnetic field\end{tabular} & 1~420~405~753(8)  & 5.60 $\times 10^{-9}$                                                                                 & 1 \\ \hline\noalign{\smallskip}

\end{tabular}
\end{table}


\section{Summary and outlook}

The ground-state hyperfine splitting of antihydrogen offers one of the most sensitive tests of CPT symmetry. ASACUSA is preparing an antihydrogen beam for an in-flight measurement of this quantity using a Rabi-type spectroscopy method which has the advantage of being performed in a field-free environment. The rate of antihydrogen formation in the CUSP needs further improvement of 1--2 orders of magnitude to allow for spectroscopy measurements.

In parallel in-beam hyperfine spectroscopy of ordinary hydrogen is pursued. A first measurement using the same cavity and sextupole as will be employed for antihydrogen resulted in a 2.7 ppb precision showing that the spectroscopy method is fully functional. Using the hydrogen beam, measurements are under way to determine  coefficients of the non-minimal Standard Model Extension that have so far not been experimentally constrained.

%

\begin{acknowledgements}
This work has been supported by the European Research Council under European Union's Seventh Framework Programme (FP7/2007-2013)/ERC Grant agreement (291242), the Austrian Ministry of Science and Research, the Austrian Science Fund (FWF): W1252-N27, a Grant-in-Aid for Specially Promoted Research (24000008) of MEXT and the RIKEN Pioneering Project. We express our gratitude towards the AD group of CERN.

\end{acknowledgements}


\clearpage\newpage

\begin{thebibliography}{10}
\providecommand{\url}[1]{{#1}}
\providecommand{\urlprefix}{URL }
\expandafter\ifx\csname urlstyle\endcsname\relax
  \providecommand{\doi}[1]{DOI~\discretionary{}{}{}#1}\else
  \providecommand{\doi}{DOI~\discretionary{}{}{}\begingroup
  \urlstyle{rm}\Url}\fi

\bibitem{TanabashiEtAl2018}
Tanabashi, M., {et al.}, {(Particle Data Group)}: Review of particle
  properties.
\newblock Phys. Rev. D \textbf{98}, 010001 (2018)

\bibitem{Ahmadi2017}
Ahmadi, M., Alves, B.X.R., Baker, C.J., Bertsche, W., Butler, E., Capra, A.,
  Carruth, C., Cesar, C.L., Charlton, M., Cohen, S., Collister, R., Eriksson,
  S., Evans, A., Evetts, N., Fajans, J., Friesen, T., Fujiwara, M.C., Gill,
  D.R., Gutierrez, A., Hangst, J.S., Hardy, W.N., Hayden, M.E., Isaac, C.A.,
  Ishida, A., Johnson, M.A., Jones, S.A., Jonsell, S., Kurchaninov, L., Madsen,
  N., Mathers, M., Maxwell, D., McKenna, J.T.K., Menary, S., Michan, J.M.,
  Momose, T., Munich, J.J., Nolan, P., Olchanski, K., Olin, A., Pusa, P.,
  Rasmussen, C.Ã., Robicheaux, F., Sacramento, R.L., Sameed, M., Sarid, E.,
  Silveira, D.M., Stracka, S., Stutter, G., So, C., Tharp, T.D., Thompson,
  J.E., Thompson, R.I., van~der Werf, D.P., Wurtele, J.S.: Observation of the
  hyperfine spectrum of antihydrogen.
\newblock Nature \textbf{548}, 66 (2017).
\newblock \urlprefix\url{http://dx.doi.org/10.1038/nature23446}

\bibitem{Ahmadi2018}
Ahmadi, M., Alves, B.X.R., Baker, C.J., Bertsche, W., Capra, A., Carruth, C.,
  Cesar, C.L., Charlton, M., Cohen, S., Collister, R., Eriksson, S., Evans, A.,
  Evetts, N., Fajans, J., Friesen, T., Fujiwara, M.C., Gill, D.R., Hangst,
  J.S., Hardy, W.N., Hayden, M.E., Isaac, C.A., Johnson, M.A., Jones, J.M.,
  Jones, S.A., Jonsell, S., Khramov, A., Knapp, P., Kurchaninov, L., Madsen,
  N., Maxwell, D., McKenna, J.T.K., Menary, S., Momose, T., Munich, J.J.,
  Olchanski, K., Olin, A., Pusa, P., Rasmussen, C.Ã., Robicheaux, F.,
  Sacramento, R.L., Sameed, M., Sarid, E., Silveira, D.M., Stutter, G., So, C.,
  Tharp, T.D., Thompson, R.I., van~der Werf, D.P., Wurtele, J.S.:
  Characterization of the 1s-2s transition in antihydrogen.
\newblock Nature \textbf{557}(7703), 71--75 (2018).
\newblock \urlprefix\url{https://doi.org/10.1038/s41586-018-0017-2}



\bibitem{Crivelli2016}
Crivelli, P., Cooke, D., Heiss, M.W.: Antiproton charge radius.
\newblock Phys. Rev. D \textbf{94}, 052008 (2016).
\newblock \doi{10.1103/PhysRevD.94.052008}.
\newblock \urlprefix\url{https://link.aps.org/doi/10.1103/PhysRevD.94.052008}

\bibitem{Ulmer:2015}
Ulmer, S., Smorra, C., Mooser, A., Franke, K., Nagahama, H., Schneider, G.,
  Higuchi, T., Van~Gorp, S., Blaum, K., Matsuda, Y., et~al.: High-precision
  comparison of the antiproton-to-proton charge-to-mass ratio.
\newblock Nature \textbf{524}(7564), 196--199 (2015)

\bibitem{Smorra2017}
Smorra, C., Sellner, S., Borchert, M.J., Harrington, J.A., Higuchi, T.,
  Nagahama, H., Tanaka, T., Mooser, A., Schneider, G., Bohman, M., Blaum, K.,
  Matsuda, Y., Ospelkaus, C., Quint, W., Walz, J., Yamazaki, Y., Ulmer, S.: A
  parts-per-billion measurement of the antiproton magnetic moment.
\newblock Nature \textbf{550}, 371 (2017).
\newblock \urlprefix\url{http://dx.doi.org/10.1038/nature24048}

\bibitem{Colladay:1997vn}
Colladay, D., Kosteleck{\'y}, V.A.: {CPT} violation and the standard model.
\newblock Physical Review D \textbf{55}, 6760--6774 (1997)

\bibitem{Kostelecky:2011jr}
Kosteleck{\'y}, V., Russell, N.: {Data tables for Lorentz and CPT violation}.
\newblock Review of Modern Physics \textbf{83}(1), 11--32 (2011)

\bibitem{Kostelecky:2018}
Kostelecky, A., Russell, N.: {Data Tables for Lorentz and CPT Violation}.
\newblock arXiv:0801.0287 (2018 edition)


\bibitem{CharltonEtAl1994}
Charlton, M., Eades, J., Horvath, D., Hughes, R., Zimmermann, C.: Antihydrogen
  physics.
\newblock Physics Reports \textbf{241}(2), 65--117 (1994)

\bibitem{HORI2013206}
Hori, M., Walz, J.: Physics at CERN's antiproton decelerator.
\newblock Progress in Particle and Nuclear Physics \textbf{72}, 206 -- 253
  (2013).
\newblock \doi{https://doi.org/10.1016/j.ppnp.2013.02.004}.
\newblock
  \urlprefix\url{http://www.sciencedirect.com/science/article/pii/S0146641013000069}

\bibitem{Ahmadi2018a}
Ahmadi, M., Alves, B.X.R., Baker, C.J., Bertsche, W., Capra, A., Carruth, C.,
  Cesar, C.L., Charlton, M., Cohen, S., Collister, R., Eriksson, S., Evans, A.,
  Evetts, N., Fajans, J., Friesen, T., Fujiwara, M.C., Gill, D.R., Hangst,
  J.S., Hardy, W.N., Hayden, M.E., Hunter, E.D., Isaac, C.A., Johnson, M.A.,
  Jones, J.M., Jones, S.A., Jonsell, S., Khramov, A., Knapp, P., Kurchaninov,
  L., Madsen, N., Maxwell, D., McKenna, J.T.K., Menary, S., Michan, J.M.,
  Momose, T., Munich, J.J., Olchanski, K., Olin, A., Pusa, P., Rasmussen, C.Ã.,
  Robicheaux, F., Sacramento, R.L., Sameed, M., Sarid, E., Silveira, D.M.,
  Starko, D.M., Stutter, G., So, C., Tharp, T.D., Thompson, R.I., van~der Werf,
  D.P., Wurtele, J.S.: Observation of the 1s-2p Lyman-$\alpha$ transition in
  antihydrogen.
\newblock Nature  \textbf{561}, 211  (2018).
\newblock \urlprefix\url{https://doi.org/10.1038/s41586-018-0435-1}

\bibitem{Widmann:2001fk}
Widmann, E., Eades, J., Hayano, R.S., Hori, M., Horv{\'a}th, D., Ishikawa, T.,
  Juh{\'a}sz, B., Sakaguchi, J., Torii, H.A., Yamaguchi, H., Yamazaki, T.:
  Hyperfine structure measurements of antiprotonic helium and antihydrogen.
\newblock In: S.G. Karshenboim, F.S. Pavone, F.~Bassani, M.~Inguscio, T.W.
  Hansch (eds.) The Hydrogen Atom: Precision Physics of Simple Atomic Systems,
  \emph{Lecture Notes in Physics}, vol. 570, pp. 528--542. Springer-Verlag
  Berlin Heidelberg (2001).
\newblock {\em arXiv:nucl-ex/0102002}

\bibitem{Widmann:2013qy}
Widmann, E., Diermaier, M., Juh{\'a}sz, B., Malbrunot, C., Massiczek, O.,
  Sauerzopf, C., Suzuki, K., W{\"u}nschek, B., Zmeskal, J., Federmann, S.,
  Kuroda, N., Ulmer, S., Yamazaki, Y.: Measurement of the hyperfine structure
  of antihydrogen in a beam.
\newblock Hyperfine Interactions \textbf{215}(1-3), 1--8 (2013).
\newblock \doi{10.1007/s10751-013-0809-6}.
\newblock \urlprefix\url{http://dx.doi.org/10.1007/s10751-013-0809-6}.
\newblock {arXiv:1301.4670}

\bibitem{Rabi:1938tq}
Rabi, I.I., Zacharias, J.R., Millman, S., Kusch, P.: {A New Method of Measuring
  Nuclear Magnetic Moment$^{ }$}.
\newblock Physical Review \textbf{53}, 318 (1938)


\bibitem{doi:10.7566/JPSCP.18.011009}
Kuroda, N., Tajima, M., Radics, B., Dupr\'e, P., Nagata, Y., Kaga, C., Kanai,
  Y., Leali, M., Rizzini, E.L., Mascagna, V., Matsudate, T., Breuker, H.,
  Higaki, H., Matsuda, Y., Ulmer, S., Venturelli, L., Yamazaki, Y.:
  Antihydrogen Synthesis in a Double-Cusp Trap (2017).
\newblock \doi{10.7566/JPSCP.18.011009}.
\newblock
  \urlprefix\url{https://journals.jps.jp/doi/abs/10.7566/JPSCP.18.011009}

\bibitem{Kuroda:2005ik}
Kuroda, N., Torii, H., Franzen, K., Wang, Z., Yoneda, S., Inoue, M., Hori, M.,
  Juh{\'a}sz, B., Horv{\'a}th, D., Higaki, H., Mohri, A., Eades, J., Komaki,
  K., Yamazaki, Y.: {Confinement of a Large Number of Antiprotons and
  Production of an Ultraslow Antiproton Beam}.
\newblock Physical Review Letters \textbf{94}(2), 023401 (2005)

\bibitem{Mohri:2003wu}
Mohri, A., Yamazaki, Y.: {A possible new scheme to synthesize antihydrogen and
  to prepare a polarised antihydrogen beam}.
\newblock Europhysics Letters \textbf{63}, 207 (2003)

\bibitem{Gabrielse1988}
Gabrielse, G., Rolston, S., Haarsma, L., Kells, W.: Antihydrogen production
  using trapped plasmas.
\newblock Physics Letters A \textbf{129}(1), 38 -- 42 (1988).
\newblock \doi{https://doi.org/10.1016/0375-9601(88)90470-7}.
\newblock
  \urlprefix\url{http://www.sciencedirect.com/science/article/pii/0375960188904707}

\bibitem{Enomoto:2010uq}
Enomoto, Y., Kuroda, N., Michishio, K., Kim, C.H., Higaki, H., Nagata, Y.,
  Kanai, Y., Torii, H.A., Corradini, M., Leali, M., Lodi-Rizzini, E., Mascagna,
  V., Venturelli, L., Zurlo, N., Fujii, K., Ohtsuka, M., Tanaka, K., Imao, H.,
  Nagashima, Y., Matsuda, Y., Juh\'asz, B., Mohri, A., Yamazaki, Y.: Synthesis
  of cold antihydrogen in a cusp trap.
\newblock Phys. Rev. Lett. \textbf{105}(24), 243401 (2010).
\newblock \doi{10.1103/PhysRevLett.105.243401}

\bibitem{Kuroda:2014fk}
Kuroda, N., Ulmer, S., Murtagh, D.J., Van~Gorp, S., Nagata, Y., Diermaier, M.,
  Federmann, S., Leali, M., Malbrunot, C., Mascagna, V., Massiczek, O.,
  Michishio, K., Mizutani, T., Mohri, A., Nagahama, H., Ohtsuka, M., Radics,
  B., Sakurai, S., Sauerzopf, C., Suzuki, K., Tajima, M., Torii, H.A.,
  Venturelli, L., W{\"u}nschek, B., Zmeskal, J., Zurlo, N., Higaki, H., Kanai,
  Y., Lodi~Rizzini, E., Nagashima, Y., Matsuda, Y., Widmann, E., Yamazaki, Y.:
  A source of antihydrogen for in-flight hyperfine spectroscopy.
\newblock Nat Commun \textbf{5}, 3089 (2014).
\newblock \urlprefix\url{http://dx.doi.org/10.1038/ncomms4089}

\bibitem{Kolbinger2018}
Kolbinger, B., Amsler, C., Breuker, H., Diermaier, M., Dupr{\'e}, P., Fleck,
  M., Gligorova, A., Higaki, H., Kanai, Y., Kobayashi, T., Leali, M.,
  M{\"a}ckel, V., Malbrunot, C., Mascagna, V., Massiczek, O., Matsuda, Y.,
  Murtagh, D., Nagata, Y., Sauerzopf, C., Simon, M.C., Tajima, M., Ulmer, S.,
  Kuroda, N., Venturelli, L., Widmann, E., Yamazaki, Y., Zmeskal, J.: Recent
  developments from ASACUSA on antihydrogen detection.
\newblock EPJ Web of Conferences \textbf{181}, 01003 (2018).
\newblock \urlprefix\url{https://doi.org/10.1051/epjconf/201818101003}

\bibitem{Malbrunot2018}
Malbrunot, C., Amsler, C., Arguedas~Cuendis, S., Breuker, H., Dupr\'e, P., Fleck,
  M., Higaki, H., Kanai, Y., Kolbinger, B., Kuroda, N., Leali, M., M{\"a}ckel,
  V., Mascagna, V., Massiczek, O., Matsuda, Y., Nagata, Y., Simon, M.C.,
  Spitzer, H., Tajima, M., Ulmer, S., Venturelli, L., Widmann, E., Wiesinger,
  M., Yamazaki, Y., Zmeskal, J.: The ASACUSA antihydrogen and hydrogen program:
  results and prospects.
\newblock Philosophical Transactions of the Royal Society of London A:
  Mathematical, Physical and Engineering Sciences \textbf{376}(2116), 20170273
  (2018).
\newblock \doi{10.1098/rsta.2017.0273}.
\newblock
  \urlprefix\url{http://rsta.royalsocietypublishing.org/content/376/2116/20170273}

\bibitem{Diermaier2017}
Diermaier, M., Jepsen, C.B., Kolbinger, B., Malbrunot, C., Massiczek, O.,
  Sauerzopf, C., Simon, M.C., Zmeskal, J., Widmann, E.: In-beam measurement of
  the hydrogen hyperfine splitting and prospects for antihydrogen spectroscopy.
\newblock Nature Communications \textbf{8}, 15749 (2017).
\newblock \urlprefix\url{http://dx.doi.org/10.1038/ncomms15749}

\bibitem{Hellwig:1970}
Hellwig, H., Vessot, R.F., Levine, M.W., Zitzewitz, P.W., Allan, D.W., Glaze,
  D.J.: Measurement of the unperturbed hydrogen hyperfine transition frequency.
\newblock Instrumentation and Measurement, IEEE Transactions on \textbf{19}(4),
  200--209 (1970)



\bibitem{Karshenboim2000}
Karshenboim, S.G.: Some possibilities for laboratory searches for variations of
  fundamental constants.
\newblock Canadian Journal of Physics \textbf{78}(7), 639--678 (2000)

\bibitem{KosteleckyVargas2015}
Kosteleck\'y, V.A., Vargas, A.J.: Lorentz and CPT tests with hydrogen,
  antihydrogen, and related systems.
\newblock Phys. Rev. D \textbf{92}, 056002 (2015).
\newblock \doi{10.1103/PhysRevD.92.056002}.
\newblock \urlprefix\url{http://link.aps.org/doi/10.1103/PhysRevD.92.056002}


\bibitem{McKeehan1936}
McKeehan, L.W.: Combinations of circular currents for producing uniform
  magnetic field gradients.
\newblock Review of Scientific Instruments \textbf{7}(4), 178--179 (1936).
\newblock \doi{10.1063/1.1752110}.
\newblock \urlprefix\url{https://doi.org/10.1063/1.1752110}



\bibitem{ArguedasCuendis2017}
Arguedas~Cuendis, S.: Measuring the hydrogen ground-state hyperfine splitting
  through the $\pi_1$ and $\sigma_1$ transitions.
\newblock M. thesis, Universit\"at Wien. Fakult\"at f\"ur Physik (2017).
\newblock \urlprefix\url{http://othes.univie.ac.at/22584/}


\bibitem{Wiesinger2017}
Wiesinger, M.: Design and implementation of new optics for the atomic hydrogen
  beam of ASACUSA's antihydrogen hyperfine spectrscopy experiment.
\newblock M. thesis, Technische Universit\"at Wien (2017).
\newblock \urlprefix\url{http://othes.univie.ac.at/22584/}




\bibitem{BreitEtAl1931}
Breit, G., Rabi, I.I.: Measurement of nuclear spin.
\newblock Phys. Rev. \textbf{38}, 2082--2083 (1931).
\newblock \doi{10.1103/PhysRev.38.2082.2}.
\newblock \urlprefix\url{https://link.aps.org/doi/10.1103/PhysRev.38.2082.2}



\bibitem{KolbingerEtAl2015}
Kolbinger, B., Capon, A., Diermaier, M., Lehner, S., Malbrunot, C., Massiczek,
  O., Sauerzopf, C., Simon, M.C., Widmann, E.: Numerical simulations of
  hyperfine transitions of antihydrogen.
\newblock Hyperfine Interactions \textbf{233}(1), 47--51 (2015).
\newblock \doi{10.1007/s10751-015-1130-3}.
\newblock \urlprefix\url{http://dx.doi.org/10.1007/s10751-015-1130-3}



\end{thebibliography}

%

\end{document}